# Charged fundamental particles in the Weyl-Dirac version of Wesson's IMT

Mark Israelit[1]

*In the framework of the Weyl-Dirac version of Wesson's Induced Matter Theory, spherically symmetric entities filled with an electrically charged substance are built in the empty 4D space-time. The substance is induced by the 5D bulk. The interior substance is characterized by density, charge density and pressure and is separated from the surrounding vacuum by a boundary surface $r_b$ where the components of the 4D metric tensor $h_{00} = 1/h_{11} = 0$. Outside of the particle one has the 4D Reissner-Nordstrøm metric with M=Q. These entities may be regarded as classical (non-quantum) models of fundamental charged particles of radius $r_b$, mass M and charge Q . Together with the neutral particle presented in a previous paper [10] we have a set of 3 fundamental particles, which are to be regarded as constituents of elementary particles (like quarks and leptons).*

---

[1] Department of Physics and Mathematics, University of Haifa-Oranim, Tivon 36006 ISRAEL
   E-mail: <israelit@macam.ac.il>

## 1. INTRODUCTION

Matter and field are basic concepts of classical field theories. They play a fundamental role in the general relativity theory [1], where the Einstein tensor $G_\mu^\nu$ is expressed in terms of the geometry of space-time, and the matter is represented by its momentum-energy density tensor $T_\mu^\nu$. These two intrinsic concepts are connected by the Einstein field equation

$$G_\mu^\nu = -8\pi\, T_\mu^\nu \ . \tag{1}$$

According to EQ. (1), a given distribution of matter (-sources) determines the geometric properties of space-time. One can regard this as the creation of space-time geometry by matter. Now, one can read EQ. (1) in the opposite direction, and expect for the creation of matter by geometry. Thus, what geometry and which mechanism have brought matter into being in our 4-dimensional world? Among others theories Wesson's Induced Matter Theory (IMT) [2, 3, 4, 5, 6, 7.] provides an elegant answer based on the creation of matter by geometry of the 5-dimensional bulk. In the Weyl-Dirac modification [8, 9] of Wesson's IMT the bulk induces on the 4D brane both, gravitation and electromagnetism, as well gravitational matter and electric current. Now, as a considerable amount of conventional matter appears in the form of particles, it would be interesting to look for a mechanism of creating fundamental particles in the framework of the Weyl-Dirac modification of Wesson's IMT.

In a recent paper [10], classical 4D models of neutral fundamental particles were considered.

In the present note we investigate the possibility of creation 4D electrically charged particles, induced by the 5D bulk in the framework of the Weyl-Dirac modification of Wesson's theory.

These particles presented in [10] and in the present paper are to be regarded as constituents of elementary particles (like quarks and leptons) and are characterized by their charge being $0;\ \pm\frac{1}{3}e$, with $e$ - the electron charge, as well by mass. Thus every quark or lepton is made up of three of these particles.

These fundamental classical particles having charge and mass are taken to be spinless and to have spherical symmetry. It is expected that, when they are quantized, they will acquire a spin, as in the case of a point particle described by the Dirac equations. Presumably the particles have other properties such as color hypercolor etc. However, these will be left to be dealt with in the future.

**In the present work following conventions are valid**: Uppercase Latin indices run from 0 to 4; lowercase Greek indices run from 0 to 3. Partial differentiation is denoted by a comma (,), Riemannian covariant 4D differentiation by a semicolon (;), and Riemannian covariant 5D differentiation by a colon (:). Further, the 5D metric tensor is denoted by $g_{AB}$, its 4D counterpart by $h_{\mu\nu}$; sometimes 5D quantities will be marked by a tilde, so $R^1_2$ is the component of the 4D Ricci tensor, whereas $\tilde{R}^1_2$ belongs to the 5D one, $R \equiv R^\sigma_\sigma$ is the 4D curvature scalar, $\tilde{R} \equiv \tilde{R}^S_S$ - the 5D one.

## 2. THE EMBEDDING FORMALISM. THE FIELD EQUATIONS

Following the ideas of Weyl [11, 12] and Dirac [13], developed by Nathan Rosen [14] and the present writer [15,16], the Weyl-Dirac version of Wesson's IMT was proposed recently [8,9]. In that version the 5D manifold {M} (the bulk) is mapped by coordinates $\{x^N\}$ and in every point exist the symmetric metric tensor $g_{AB}$, as well the Weylian connection vector $\tilde{w}_C$ and the Dirac gauge function $\Omega$. The three fields $g_{AB}$, $\tilde{w}_C$ and $\Omega$ are integral parts of the geometric framework, and no additional fields, sources or particles exist in the bulk {M}. In this 5D manifold, field equations for $g_{AB}$ and $\tilde{w}_C$, are derived from a geometrically based action. It turns out that the equation for $\Omega$ is actually a corollary of the $g_{AB}$, and $\tilde{w}_C$ equations, so that the Dirac gauge function may be chosen arbitrarily.

Below follows a concise description of the general embedding formalism. The notations as well as the geometric construction given below accord to these given in works of Paul Wesson and Sanjeev S. Seahra [2, 3, 4, 5, 6], as well in works of the present writer [8, 9, 10].

One considers a 5-dimensional manifold { M } (the "bulk") with a symmetric metric $g_{AB} = g_{BA}$, having the signature $\text{sig}(g_{AB}) = (+,-,-,-,\varepsilon)$ with $\varepsilon = \pm 1$. The manifold is mapped by coordinates $\{x^A\}$ and described by the line-element

$$dS^2 = g_{AB} dx^A dx^B \quad (A, B = 0, 1, 2, 3, 4) \qquad (2)$$

One can introduce a scalar function $l = l(x^A)$ that defines the foliation of {M} with 4-dimensional hyper-surfaces $\Sigma_l$ at a chosen $l$ = const, as well the vector $n^A$ normal to $\Sigma_l$. If there is only one time-like direction in {M}, it will be assumed that $n^A$ is space-like. If $\{M\}$ possesses two time-like directions $(\varepsilon = +1)$, $n^A$ is a time-like vector. Thus, in any case $\Sigma_l$ (the "brane") contains three space-like directions and a time-like one. The brane, our 4-dimensional space-time, is mapped by coordinates $\{y^\mu\}$, and has the metric $h_{\mu\nu} = h_{\nu\mu}$ with $\text{sig}(h_{\mu\nu}) = (+,-,-,-)$. The line-element on the brane is (cf. (2))

$$ds^2 = h_{\mu\nu} dy^\mu dy^\nu \qquad (\mu,\nu = 0,1,2,3) \quad (3)$$

It is supposed that the relations $y^\nu = y^\nu(x^A)$ and $l = l(x^A)$, as well as the reciprocal one $x^A = x^A(y^\nu, l)$ are mathematically well-behaved functions. Thus, the 5D bulk may be mapped either by $\{x^A\}$ or by $\{y^\nu, l\}$.

A given 5D quantity (vector, tensor) in the bulk has a 4D counterpart located on the brane. These counterparts may be formed by means of the following system of basis vectors, which are orthogonal to $n_A$

$$e_\nu^A = \frac{\partial x^A}{\partial y^\nu} \qquad \text{with} \qquad n_A e_\nu^A = 0 \qquad (4)$$

The brane $\Sigma_l$ is stretched on four $(\nu = 0,1,2,3)$ five-dimensional basis vectors $e_\nu^A$. In addition to the main basis $\{e_\nu^A; n_A\}$ one can consider its associated one $\{e_A^\nu; n^A\}$, which also satisfies the orthogonality condition $e_A^\nu n^A = 0$. The main basis and its associated are connected by the following relations:

$$e_\nu^A e_A^\mu = \delta_\nu^\mu \;;\quad e_\sigma^A e_B^\sigma = \delta_B^A - \varepsilon n^A n_B \;;\quad n^A n_A = \varepsilon \qquad (5)$$

Let us consider a 5D vector $V_A ; V^A$ in the bulk $\{M\}$. Its 4D counterpart on the brane $\Sigma_l$ is given by

$$V_\mu = e_\mu^A V_A \;;\quad V^\nu = e_B^\nu V^B . \qquad (6)$$

On the other hand the 5D vector may be written as

$$V_A = e_A^\mu V_\mu + \varepsilon(V_S n^S) n_A ; \quad V^A = e_\mu^A V^\mu + \varepsilon(V^S n_S) n^A \tag{7}$$

Further, the 5D metric tensor, $g_{AB}$; $g^{AB}$, and the 4D one, $h_{\mu\nu}$; $h^{\mu\nu}$, are related by

$$h_{\mu\nu} = e_\mu^A e_\nu^B g_{AB} ; \quad h^{\mu\nu} = e_A^\mu e_B^\nu g^{AB} ; \quad \text{with} \quad h_{\mu\nu} h^{\lambda\nu} = \delta_\mu^\lambda \tag{8}$$

$$g_{AB} = e_A^\mu e_B^\nu h_{\mu\nu} + \varepsilon n_A n_B ; \quad g^{AB} = e_\mu^A e_\nu^B h^{\mu\nu} + \varepsilon n^A n^B ; \quad \text{with} \quad g_{AB} g^{CB} = \delta_A^C \tag{9}$$

Considering the bulk of the Weyl-Dirac modification of Wesson's IMT we have to pay attention to the Weylian connection vector $\tilde{w}_A$ and to the 5D field tensor $\tilde{W}_{AB} \equiv \tilde{w}_{A,B} - \tilde{w}_{B,A}$. There is also the Dirac gauge function $\Omega(x^B)$ and its partial derivative $\Omega_A \equiv \dfrac{\partial \Omega}{\partial x^A}$. On the 4D brane one has the metric $h_{\mu\nu}$, the 4D Weyl vector $w_\mu$, the 4D Maxwell field tensor $W_{\mu\nu} = w_{\mu,\nu} - w_{\nu,\mu}$ and the gauge function.

Starting from the 5D equations for the metric $g_{AB}$ and making use of the Gauss-Codazzi equations the following 4-D equations of gravitation may be derived [8, 9]:

$$G_{\alpha\beta} = -\frac{8\pi}{\Omega^2} M_{\alpha\beta} - \frac{2\varepsilon}{\Omega^2}\left(\frac{1}{2}h_{\alpha\beta}B - B_{\alpha\beta}\right) + \frac{6}{\Omega^2}\Omega_\alpha \Omega_\beta - \frac{3}{\Omega}\left(\Omega_{\alpha;\beta} - h_{\alpha\beta}\Omega^\sigma_{;\sigma}\right)$$
$$+\frac{3\varepsilon}{\Omega}(\Omega_S n^S)(h_{\alpha\beta}C - C_{\alpha\beta}) + \varepsilon\left[E_{\alpha\beta} - h_{\alpha\beta}E + h^{\mu\nu}C_{\mu[\nu}C_{\lambda]\sigma}(h_{\alpha\beta}h^{\lambda\sigma} - 2\delta_\alpha^\sigma \delta_\beta^\lambda)\right] - \frac{1}{2}h_{\alpha\beta}\Omega^2 \Lambda \tag{10}$$

From the equation of the **source-free** 5D Weylian field $(\Omega \tilde{W}^{AB})_{:B} = 0$ in the bulk, was derived in [8] the 4D equation for the Maxwell field $W_{\mu\nu}$ on the brane

$$W^{\alpha\beta}_{;\beta} = -\frac{\Omega_\beta}{\Omega} W^{\alpha\beta} + \varepsilon n_S \left[\tilde{W}^{AS}(e_A^\beta h^{\alpha\lambda} - e_A^\alpha h^{\beta\lambda})C_{\beta\lambda} + n^C e_A^\alpha \left(\tilde{W}^{AS}_{:C} + \tilde{W}^{AS}\frac{\Omega_C}{\Omega}\right)\right] \tag{11}$$

In (10) and (11) the following quantities appear:

a) The conventional energy-momentum density tensor of the 4D electromagnetic field

$$M_{\alpha\beta} = \frac{1}{4\pi}\left(\frac{1}{4}h_{\alpha\beta}W_{\lambda\sigma}W^{\lambda\sigma} - W_{\alpha\lambda}W_{\beta}^{\ \cdot\lambda}\right) \tag{12a}$$

b) Energy-momentum quantities formed from the 5D Weylian field $\tilde{W}_{AB}$ (cf. [8])

$$B_{\alpha\beta} \equiv \tilde{W}_{AS}\tilde{W}_{BL} e_\alpha^A e_\beta^B n^S n^L \quad \text{and} \quad B = h^{\lambda\sigma}B_{\lambda\sigma} \equiv \tilde{W}_{AS}\tilde{W}_{BL} g^{AB} n^S n^L \tag{12b}$$

c) The extrinsic curvature $C_{\mu\nu}$ of the brane $\Sigma_l$, and its contraction $C$

$$C_{\mu\nu} = e_\mu^A e_\nu^B n_{B:A} \equiv e_\mu^A e_\nu^B\left(\frac{\partial n_B}{\partial x^A} - n_S \tilde{\Gamma}^S_{AB}\right), \quad C = h^{\lambda\sigma}C_{\lambda\sigma} \tag{12c}$$

d) A quantity formed from the 5D curvature tensor (cf. [3, 4, 5])

$$E_{\alpha\beta} \equiv \tilde{R}_{MANB}\, n^M n^N e_\alpha^A e_\beta^B \tag{12d}$$

e) as well its contraction

$$E \equiv h^{\lambda\sigma}E_{\lambda\sigma} = -R_{MN}\, n^M n^N \tag{12e}$$

Finally, in (10, 11), $G_{\mu\nu}$ stands for the Einstein tensor, $\Lambda$ is the cosmological constant and $\Omega_{,A} \equiv \Omega_A$; $g^{AB}\Omega_{,B} \equiv \Omega^A$.

Details may be found in Ref. 8.

## 3. THE STATIC SPHERICALLY SYMMETRIC CASE

In order to describe a particle-like entity in the 4D brane, which is mapped by the coordinates $y^0 = t$; $y^1 = r$; $y^2 = \vartheta$; $y^3 = \varphi$, we write the spherically symmetric static line element as

$$ds^2 = e^{\nu(r)}dt^2 - e^{\lambda(r)}dr^2 - r^2\left(d\vartheta^2 + \sin^2\vartheta\, d\varphi^2\right) \tag{13}$$

It is believed that the entity is restricted by a boundary sphere of radius $r = r_b$; the interior ($r \leq r_b$) is filled with a substance induced by the bulk and described by matter density $\rho$, charge density $\rho_e$ and pressure $P$. These three characteristic functions have no singularity at $r = 0$ and vanish at the boundary. Outside ($r > r_b$) there is vacuum.

The 5D bulk is mapped by $x^0 = e^{-\frac{1}{2}N(l)} y^0$; $x^1 = e^{-\frac{1}{2}L(l)} y^1$; $x^2 = y^2$; $x^3 = y^3$: $x^4 = l$, (the functions $N(l), L(l)$ are defined in (15)) and the 5D line element will be written as

$$dS^2 = g_{AB} dx^A dx^B = e^{\tilde{N}(r,l)}(dx^0)^2 - e^{\tilde{L}(r,l)}(dx^1)^2 - r^2(d\vartheta^2 + \sin^2\vartheta\, d\varphi^2) + \varepsilon\, e^{\tilde{F}(r,l)} dl^2 \quad (14)$$

It is convenient to divide the metric functions into depending on $r$ and depending on $l$, writing

$$\tilde{N}(r,l) = N(l) + \nu(r); \quad \tilde{L}(r,l) = L(l) + \lambda(r); \quad \tilde{F}(r,l) = F(l) + \psi(r) \quad (15)$$

Hereafter, we denote a partial derivative with respect to $r$ by a prime and that with respect to the fifth coordinate, $l$ by a dot. Without any restriction we can impose the condition $N(l_0) = L(l_0) = F(l_0) = 0$ for the values on the brane $l = l_0$, our 4D space-time. The basic vectors, the metrics as well the Christoffel symbols of (13-15) are given by (A-1) – (A-5) in the Appendix.

Besides the metric tensor $g_{AB}$, the bulk possesses the Weyl vector $\tilde{w}_A$, which has the following components

$$\tilde{w}_0(x^1, l); \quad \tilde{w}_4(x^1, l) \quad (16)$$

From it one forms the 5D Weylian field

$$\tilde{W}_{01} = \tilde{w}_{0,1}; \quad \tilde{W}^{01} = -e^{-(\tilde{L}+\tilde{N})} \tilde{w}_{0,1}; \quad \tilde{W}_{14} = -\tilde{w}_{4,1}; \quad \tilde{W}^{14} = \varepsilon\, e^{-(\tilde{L}+\tilde{F})} \tilde{w}_{4,1};$$
$$\tilde{W}_{04} = \dot{\tilde{w}}_0; \quad \tilde{W}^{04} = \varepsilon\, e^{-(\tilde{N}+\tilde{F})} \dot{\tilde{w}}_0 \quad (17)$$

and as $N(l_0) = L(l_0) = F(l_0) = 0$ we have for the 4D Maxwell field on the brane

$$\tilde{w}'_0 = e^{\frac{1}{2}N} w'_0 \equiv w'_0; \quad W_{01} = e^{-\frac{1}{2}(N+L)} \tilde{W}_{01} = e^{-\frac{1}{2}L(l_0)} w'_0 \cong w'_0 \tag{18}$$

There is also the Dirac gauge function $\Omega$ and its partial derivative $\Omega_A \equiv \dfrac{\partial \Omega}{\partial x^A}$

We assume $\Omega = \Omega(r)$ so that

$$\Omega_A = 0 \quad \text{for} \quad A \neq 1 \tag{19}$$

It must be emphasized that the 5D bulk is empty – it possesses no matter fields. The functions $\Omega$, $\tilde{w}_0(x^1, l)$; $\tilde{w}_4(x^1, l)$ and $\tilde{W}_{AB}$ are essential parts of the 5D Weyl-Dirac geometric framework in the bulk. On the other hand their 4D counterparts $w'_0$ and $W_{01}$ are regarded as representing the Maxwell field with sources induced by the bulk.

It is convenient to write the gravitational equation (10) in its co-contravariant form. Further, we take into account that by (19) and (A-3) $\Omega_S n^S = 0$. Thus we have

$$G_\alpha^\beta = -\frac{8\pi}{\Omega^2} M_\alpha^\beta - \frac{2\varepsilon}{\Omega^2}\left(\frac{1}{2}\delta_\alpha^\beta B - B_\alpha^\beta\right) + \frac{6h^{\beta\lambda}}{\Omega^2}\Omega_\alpha \Omega_\lambda - \frac{3}{\Omega}\left(h^{\beta\lambda}\Omega_{\alpha;\lambda} - \delta_\alpha^\beta \Omega^\sigma_{;\sigma}\right) + \\ + \varepsilon\left[E_\alpha^\beta - \delta_\alpha^\beta E + h^{\mu\nu} C_{\mu[\nu} C_{\lambda]\sigma}\left(\delta_\alpha^\beta h^{\lambda\sigma} - 2\delta_\alpha^\sigma h^{\lambda\beta}\right)\right] - \frac{1}{2}\delta_\alpha^\beta \Omega^2 \Lambda \tag{20}$$

The quantities appearing in (20) and listed in (12a-12e) may be accounted making use of (16-18), as well of (A-3) - (A-5). The result is listed in (A-6)

For the sake of convenience we turn to an auxiliary gauge function $\omega(r) = \ln \Omega(r)$. Then, making use of (A-6a – A-6e) we obtain from (20) the following gravitational equations:

$$G_0^0 = -e^{-2\omega}e^{-(\lambda+\nu)}(\tilde{w}_0')^2 + \varepsilon\, e^{-(2\omega+\psi)}\left[e^{-\nu}(\dot{\tilde{w}}_0)^2 + e^{-\lambda}(\tilde{w}_{4,1})^2\right]-$$

$$-\frac{1}{2}e^{-\lambda}\left[\psi'' + \frac{1}{2}(\psi')^2 - \frac{1}{2}\lambda'\psi' + 2\frac{\psi'}{r}\right] - 3e^{-\lambda}\left[\omega'' + (\omega')^2 - \frac{1}{2}\lambda'\omega' + 2\frac{\omega'}{r}\right] + \quad (21)$$

$$+\frac{\varepsilon}{2}e^{-\psi}\left[\ddot{L} + \frac{1}{2}(\dot{L})^2 - \frac{1}{2}\dot{F}\dot{L}\right] - \frac{1}{2}e^{2\omega}\Lambda$$

$$G_1^1 = -e^{-2\omega}e^{-(\lambda+\nu)}(\tilde{w}_0')^2 - \varepsilon\, e^{-(2\omega+\psi)}\left[e^{-\nu}(\dot{\tilde{w}}_0)^2 + e^{-\lambda}(\tilde{w}_{4,1})^2\right] - \frac{1}{2}e^{-\lambda}\left(\frac{1}{2}\nu'\psi' + 2\frac{\psi'}{r}\right)$$

$$-3e^{-\lambda}\left[2(\omega')^2 + \frac{1}{2}\nu'\omega' + 2\frac{\omega'}{r}\right] + \frac{\varepsilon}{2}e^{-\psi}\left[\ddot{N} + \frac{1}{2}(\dot{N})^2 - \frac{1}{2}\dot{F}\dot{N}\right] - \frac{1}{2}e^{2\omega}\Lambda \quad (22)$$

$$G_2^2 = e^{-2\omega}e^{-(\lambda+\nu)}(\tilde{w}_0')^2 - \varepsilon\, e^{-(2\omega+\psi)}\left[e^{-\nu}(\dot{\tilde{w}}_0)^2 - e^{-\lambda}(\tilde{w}_{4,1})^2\right]$$

$$-\frac{1}{2}e^{-\lambda}\left[\psi'' + \frac{1}{2}(\psi')^2 + \frac{1}{2}\psi'(\nu'-\lambda') + \frac{\psi'}{r}\right] - 3e^{-\lambda}\left[\omega'' + (\omega')^2 + \frac{1}{2}\omega'(\nu'-\lambda') + \frac{\omega'}{r}\right] + \quad (23)$$

$$+\frac{\varepsilon}{2}e^{-\psi}\left[\ddot{N} + \ddot{L} + \frac{1}{2}(\dot{L})^2 + \frac{1}{2}(\dot{N})^2 - \frac{1}{2}\dot{F}(\dot{L}+\dot{N}) + \frac{1}{2}\dot{L}\dot{N}\right] - \frac{1}{2}e^{2\omega}\Lambda$$

In addition to (21-23) we have from (11) the Maxwell equation that takes the form

$$\frac{\partial}{\partial r}\left(e^{-\frac{1}{2}(\lambda+\nu+\psi)}r^2\Omega w_0'\right) = -\varepsilon\, e^{\frac{1}{2}(\lambda-\nu-3\psi)}\left[\ddot{\tilde{w}}_0 + \frac{1}{2}(\dot{F}+\dot{L}+\dot{N})\dot{\tilde{w}}_0\right]r^2\Omega \quad (24)$$

Integrating (24) we get

$$w_0' = -\frac{\varepsilon}{r^2}e^{\frac{1}{2}(\lambda+\nu+\psi-2\omega)}\left[\int_0^r e^{\frac{1}{2}(\lambda-\nu-3\psi+2\omega)}\left[\ddot{\tilde{w}}_0 + \frac{1}{2}(\dot{F}+\dot{L}+\dot{N})\dot{\tilde{w}}_0\right]r^2 dr + Const.\right] \quad (25)$$

In the above treatment are 4 equations (21 - 23, 25) for six functions, $\lambda, \nu, \psi, \omega, \tilde{w}_0, \tilde{w}_4$ (The quantities $\ddot{L}, \ddot{N}, \dot{L}, \dot{N}, \dot{F}$ are constants on the brane $l = l_0$). Thus, we can impose two conditions. To choose these conditions, we assume that the entity is filled with a perfect fluid, so that EQ-s (21-23) may be rewritten as

$$G_0^0 \equiv e^{-\lambda}\left(-\frac{\lambda'}{r} + \frac{1}{r^2}\right) - \frac{1}{r^2} = -\frac{\tilde{q}^2}{r^4} - 8\pi\rho \quad (21a)$$

$$G_1^1 \equiv e^{-\lambda}\left(\frac{v'}{r}+\frac{1}{r^2}\right)-\frac{1}{r^2}=-\frac{\tilde{q}^2}{r^4}+8\pi P_n \tag{22a}$$

$$G_2^2 \equiv e^{-\lambda}\left[\frac{v''}{2}-\frac{\lambda'v'}{4}+\frac{(v')^2}{4}+\frac{v'-\lambda'}{2r}\right]=\frac{\tilde{q}^2}{r^4}+8\pi P_\tau \tag{23a}$$

The quantity $\tilde{q}(r)$ is regarded as the effective charge inside a sphere of radius $r$ and according to (21) and (25)) it is given by

$$\tilde{q}=-\varepsilon\, e^{\frac{1}{2}(\psi-4\omega)}\int_0^r e^{\frac{1}{2}(\lambda-\nu-3\psi+2\omega)}\left[\ddot{\tilde{w}}_0+\frac{1}{2}\left(\dot{F}+\dot{L}+\dot{N}\right)\dot{\tilde{w}}_0\right]r^2\,dr \tag{26}$$

(We discarded the constant term in (26) as leading to a singular point charge.)

Thus, the term $\dfrac{\tilde{q}^2}{r^4}$ is the electromagnetic energy inside the sphere of radius $r$.

Further, $8\pi\rho(r)$, which includes the remaining terms in the RHS of (21), is the matter density inside the spherically symmetric entity, $P_n(r)$ is the radial pressure and $P_\tau(r)$ stands for the tangential pressure.

We are looking for a non-rotating entity filled with perfect fluid, therefore we impose

$$P_\tau = P_n = P \tag{27}$$

The second condition will be imposed in order to get the following equation of state [2]

$$P = -\rho \tag{28}$$

Condition (27) imposed on (22, 23) yields

---

[2] Following previous papers [17] we will refer to matter in such a state as "prematter" and regard it as a primary substance.

$$2\varepsilon\, e^{-(\lambda+\psi+2\omega)}\left(\tilde{w}_{4,1}\right)^2 = -\frac{\varepsilon}{2}e^{-\psi}\left[\ddot{L}+\frac{1}{2}(\dot{L})^2-\frac{1}{2}\dot{F}\dot{L}+\frac{1}{2}\dot{L}\dot{N}\right]+$$
$$+\frac{1}{2}e^{-\lambda}\left[\psi''+\frac{1}{2}(\psi')^2-\frac{1}{2}\lambda'\psi'-\frac{\psi'}{r}\right]+3e^{-\lambda}\left[\omega''-(\omega')^2-\frac{1}{2}\lambda'\omega'-\frac{\omega'}{r}\right] \quad (29)$$

Condition (28) describes prematter and it leads to

$$2\varepsilon\, e^{-(\nu+\psi+2\omega)}\left(\dot{\tilde{w}}_0\right)^2 = e^{-\lambda}\left(\frac{1}{2}\psi'+3\omega'\right)\left(\frac{1}{r}-\frac{1}{2}\nu'\right)+$$
$$+\frac{\varepsilon}{2}e^{-\psi}\left[\ddot{N}+\frac{1}{2}(\dot{N})^2-\frac{1}{2}\dot{F}\dot{N}+\frac{1}{2}\dot{L}\dot{N}\right] \quad (30)$$

A restriction $B_{01}=0$ follows from $G_{01}=0$. Thus, as $B_{01}=-e^{-\frac{1}{2}(L+N+2\tilde{F})}\dot{\tilde{w}}_0\tilde{w}_{4,1}$ (cf. (A6b)), one has two possibilities either,

$$\dot{\tilde{w}}_0 = 0 \quad (31)$$

or

$$\tilde{w}_{4,1} = 0 \quad (32)$$

Equations (21-23, 25) with conditions (29, 30) describe the spherically symmetric, static prematter entity. This system of equations, however, seems to be very much complicated; therefore we will simplify our case.

## 4. THE SIMPLIFIED CASE

Let us presume (cf. (15))

$$N(l) = L(l) \equiv 0; \quad (33)$$

Then, choosing the possibility (31), $\dot{\tilde{w}}_0 = 0$ one has according to (30)

$$\left(\frac{1}{2}\psi' + 3\omega'\right)\left(\frac{1}{r} - \frac{1}{2}\nu'\right) = 0 \tag{34}$$

This results in a very simple gauge condition

$$\omega' = -\frac{1}{6}\psi'; \quad \omega = -\frac{1}{6}\psi \tag{35}$$

(We discard a possible constant in the second relation (35).)

With (33) and (35) we have from (29)

$$2\varepsilon\, e^{-(\lambda+\psi+2\omega)}(\tilde{w}_{4,1})^2 = \frac{1}{6} e^{-\lambda}(\psi')^2 \tag{36}$$

Making use of (33), (35), (36) and discarding the cosmological term as irrelevant one obtains from (21-23) the following equations:

$$G_0^0 = -e^{-\left(\lambda+\nu-\frac{\psi}{3}\right)}(\tilde{w}_0')^2 - \frac{e^{-\lambda}(\psi')^2}{4} \tag{37a}$$

$$G_1^1 = -e^{-\left(\lambda+\nu-\frac{\psi}{3}\right)}(\tilde{w}_0')^2 - \frac{e^{-\lambda}(\psi')^2}{4} \tag{37b}$$

$$G_2^2 = e^{-\left(\lambda+\nu-\frac{\psi}{3}\right)}(\tilde{w}_0')^2 - \frac{e^{-\lambda}(\psi')^2}{4} \tag{37c}$$

From (37a) and (37b) one concludes that

$$\lambda + \nu = 0 \tag{38}$$

If one makes use of (25), (31), (33), (35), (36) and takes into account relation (38), one obtains the Maxwell EQ.

$$w_0' = -\frac{\varepsilon}{r^2} e^{\frac{2}{3}\psi} \int_0^r e^{\left(\lambda-\frac{5}{3}\psi\right)} \ddot{\tilde{w}}_0\, r^2 dr \tag{39}$$

According to (37a) and (39) we have the effective charge inside the sphere of radius $r$

$$\tilde{q}(r) = e^{\frac{5}{6}\psi} \int_0^r e^{\left(\lambda - \frac{5}{3}\psi\right)} \ddot{w}_0 \, r^2 dr \tag{40}$$

Further, from (28) and (37a-c) one has

$$8\pi\rho = -8\pi P = \frac{1}{4} e^{-\lambda} (\psi')^2 \tag{41}$$

With (40) and (41) one rewrites EQ-s (37a-c) as

$$G_0^0 \equiv e^{-\lambda}\left(-\frac{\lambda'}{r} + \frac{1}{r^2}\right) - \frac{1}{r^2} = -8\pi\rho - \frac{\tilde{q}^2}{r^4} \tag{42}$$

$$G_1^1 \equiv e^{-\lambda}\left(\frac{v'}{r} + \frac{1}{r^2}\right) - \frac{1}{r^2} = 8\pi P - \frac{\tilde{q}^2}{r^4} \tag{43}$$

$$G_2^2 \equiv \frac{e^{-\lambda}}{2}\left(v'' - \frac{\lambda' v'}{2} + \frac{(v')^2}{2} + \frac{v' - \lambda'}{r}\right) = 8\pi P + \frac{\tilde{q}^2}{r^4} \tag{44}$$

As noted above, the entity is restricted by a sphere of radius $r_b$. Inside there is the prematter substance, outside one has vacuum. Accordingly, introducing the function $y(r) \equiv e^{-\lambda} \equiv e^v$ one obtains the following solution of (42) and (43)

$$y(r) \equiv e^v \equiv e^{-\lambda} = 1 - \frac{8\pi}{r}\int_0^r \rho r^2 dr - \frac{1}{r}\int_0^r \frac{\tilde{q}^2}{r^2} dr; \quad \text{for } r \le r_b \tag{45}$$

and

$$y(r) = 1 - \frac{2M}{r} + \frac{Q^2}{r^2}; \quad \text{for } r > r_b; \text{ and with } Q \equiv \tilde{q}(r_b) \tag{46}$$

In EQ. (46) $M$ stands for the mass of the whole entity, while, according to (40) the total charge $Q$ is given by

$$Q \equiv \tilde{q}(r_b) = e^{\frac{5}{6}\psi} \int_0^{r_b} e^{\left(\lambda - \frac{5}{3}\psi\right)} \ddot{w}_0 \, r^2 dr \qquad (47)$$

From the two equations (45, 46) we obtain for the mass as seen by an external observer

$$M = \frac{Q^2}{2r_b} + 4\pi \int_0^{r_b} \rho r^2 dr + \frac{1}{2}\int_0^{r_b} \frac{\tilde{q}^2}{r^2} dr \qquad (48)$$

For a moment let us go back to the equations (42)-(44). Instead of solving (44) we can make use of the equilibrium relation

$$8\pi\rho' + 8\pi(\rho + P) = -\frac{2\tilde{q}\tilde{q}'}{r^4} = -\frac{(\tilde{q}^2)'}{r^4} \qquad (49)$$

Taking into account (28) this relation may rewritten as

$$8\pi\rho' = -\frac{(\tilde{q}^2)'}{r^4} \qquad (50)$$

so that

$$(\tilde{q})^2 = -8\pi r^4 \rho + 32\pi \int_0^r \rho r^3 dr \qquad (51)$$

However, as noted above, $\rho(r_b) = 0$. Thus, the total charge is given by

$$Q^2 = +32\pi \int_0^{r_b} \rho r^3 dr \qquad (52)$$

## 5. THE MODEL

Let us go back to (42) and substitute into it $y(r) = e^{-\lambda} = e^{\nu}$. Then we obtain

$$\frac{y'}{r} + \frac{y}{r^2} - \frac{1}{r^2} = -8\pi\rho - \frac{\tilde{q}^2}{r^4} \tag{53}$$

Making use of (51) one has

$$y'r^3 + yr^2 - r^2 = -32\pi \int_0^r \rho r^3 dr \tag{54}$$

This may be rewritten as

$$y'' + \frac{4}{r} y' + \frac{2}{r^2} y - \frac{2}{r^2} = -32\pi\rho \tag{55}$$

For $z = y - 1$ the left hand side of EQ. (55) looks like a Bessel equation. Therefore we shall look for an appropriate Bessel function, or a combination of them. There is the spherical Bessel function of the first kind $j_0(x) = \frac{\sin x}{x}$. But $j_0(x)$ can take negative values, therefore we take its square as a possible description of $y(r)$

$$y = \frac{1}{k^2 r^2} \sin^2(kr) \tag{56}$$

with $k \equiv \frac{\pi}{r_b}$ ; (note that $|k| = cm^{-1}$)

Inserting (56) into (55) one obtains

$$8\pi\rho = \frac{1 - \cos(2kr)}{2r^2} = \frac{\sin^2(kr)}{r^2} \equiv k^2 y \tag{57}$$

It is worth noting that the mass density $\rho(r) \geq 0$; $8\pi\rho(0) = k^2$; $\rho(r_b) = 0$.

Further, substituting $\rho$ into (51), and choosing a suitable value of the constant of integration, we obtain the effective charge inside the entity

$$\tilde{q}^2(r) = \left[r\cos(kr) - \frac{1}{k}\sin(kr)\right]^2; \quad \text{and} \quad \tilde{q}(r) = \pm\left[r\cos(kr) - \frac{1}{k}\sin(kr)\right] \quad (58)$$

We emphasize that according to (58)

$$\tilde{q}(0) = 0; \quad \text{and} \quad |Q| \equiv |\tilde{q}(r_b)| = r_b \quad (59)$$

To obtain $\psi$ one can equate (41) and (57). This leads to the result

$$(\psi')^2 = 4k^2 \Rightarrow \psi' = \pm 2k; \quad \text{and} \quad \psi = \pm 2k\, r + Const \quad (60)$$

Choosing $Const = \mp 2\pi$ we have

$$\psi = \pm(2k\, r - 2\pi) \quad (60a)$$

so that $\psi(r=0) = \mp 2\pi$ and $\psi(r=r_b) = 0$. We will also assume $\psi = 0$ for $r > r_b$.

To account the external mass $M$, one starts from (48) and makes use of (57) and (58). As a result one obtains

$$M = \frac{Q^2}{2r_b} + \frac{1}{2}r_b \quad (61)$$

and making use of (59) one has

$$M = Q = r_b$$

It is interesting that for neutral particles (cf. ref. (10)) we obtained $M_{neutral} = \frac{1}{2}r_b$.

In order to obtain the charge density $\rho_e$ inside the entity we recall that for a spherically symmetric distribution of matter the charge is given by $q = 4\pi \int_0^r e^{\frac{\lambda}{2}} \rho_e\, r^2 dr$. Making use of (58) and (56) one obtains

$$4\pi|\rho_e| = \frac{\sin^2 kr}{r^2} \qquad (62)$$

Thus, according to (57)

$$|\rho_e| = 2\rho \qquad (62a)$$

It would be of course interesting to obtain the function $\ddot{\tilde{w}}_0$, which invoked the charge. According to EQ. (40) we have

$$\tilde{q}' = \frac{5}{6}\psi'\tilde{q} + e^{\lambda - \frac{5}{6}\psi} \ddot{\tilde{w}}_0 r^2 \qquad (63)$$

After elementary calculation with $|\tilde{q}|$ from EQ. (58) we get

$$\ddot{\tilde{w}}_0 = e^{\frac{5}{3}(kr-\pi)} \left[ \frac{5}{3}\frac{\sin(kr)}{r^2} - \frac{5}{3}\frac{k\cos(kr)}{r} - k\frac{\sin(kr)}{r} \right] \frac{\sin^2(kr)}{k^2 r^2} \qquad (64)$$

Expanding the two first terms in the bracket for small values of $kr$ we obtain

$$\left|\ddot{\tilde{w}}_0(0)\right| = e^{-\frac{5}{3}\pi} k^2 \equiv e^{-\frac{5}{3}\pi} \frac{\pi^2}{r_b^2} \quad, \text{ whereas } \quad \ddot{\tilde{w}}_0(r_b) = 0; \qquad (65)$$

We see that there is no singularity at the center, whereas at the boundary $\ddot{\tilde{w}}_0$ vanishes.

Now, with $\psi$ given in (60a) we can account the strength of the Maxwell field $w_0'$ according to (39)

$$w_0' = -\frac{\varepsilon}{r^2} e^{\pm\frac{1}{3}(\pi-kr)} \tilde{q}(r) \quad \text{for} \quad r \leq r_b \qquad (66)$$

$$w_0' = -\frac{\varepsilon}{r^2} Q \qquad \text{for} \quad r > r_b \qquad (67)$$

We obtained a plausible model, describing 4D fundamental particles created by the bulk in the Weyl-Dirac modification of Wesson's IMT.

Besides the model considered above there are of course other ones; so one could consider the case with $\tilde{w}_{4,1} = 0$, but $\dot{\tilde{w}}_0 \neq 0$, or the entity in the Einstein gauge and more models.

In the above procedure we obtained relations for some 4D quantities that originate from the bulk. These results are somehow incomplete. So, in (64) we found, rather $\ddot{\tilde{w}}_0(r,l_0)$, than $\ddot{\tilde{w}}_0(r,l)$ for any $l$; in (32a) we assumed $\dot{\tilde{w}}_0(r,l_0) = 0$, but $\dot{\tilde{w}}_0(r,l) \neq 0$; in equation (36) appears $\tilde{w}_{4,1}(r,l_0)$ i.e. again the value on the brane. Finally $F(l)$ remains arbitrary.

## 6. CONCLUSIONS AND DISCUSSION

The aim of the present paper is to investigate the possibility of creating charged fundamental particles in our 4D space-time regarded as a brane in the 5D manifold, the latter being the bulk of the Weyl-Dirac modification of Wesson's Induced Matter Theory. This bulk is an empty (without matter, charges, currents) 5D Weyl-Dirac manifold described by the metric tensor $g_{AB}$, the Weyl connection vector $\tilde{w}_A$ as well by the Dirac gauge function $\Omega$.

As shown previously the bulk creates-induces matter [1-7], as well electric charges, currents and the Maxwell field [8, 9] in the brane. Recently [10] the creation of neutral particles in the Weyl-Dirac modification of Wesson's IMT was considered. In this paper we consider a spherically symmetric metric in the bulk and a spherically symmetric entity filled with induced charged matter in the brane. This entity is restricted by a border surface of radius $r_b$ so that beyond it one has vacuum.

A special, very interesting analytical solution for a plausible model was found. In it, the substance filling the interior is characterized by a mass density $\rho(r)$ and by a charge density $\rho_e = 2\rho$, both vanishing at the border $r = r_b$. In the center one has $8\pi\rho(0) = k^2$ with $k \equiv \frac{\pi}{r_b}$. In the interior acts the electric field given by $w_0' = -\frac{\varepsilon}{r^2} e^{\frac{1}{3}(\pi - kr)} \tilde{q}(r)$ with $\tilde{q}(r)$ being the charge inside the sphere of radius $r$, whereas for $r > r_b$ one has $w_0' = -\frac{\varepsilon}{r^2} Q$ with $Q = \tilde{q}(r_b)$. The metric inside the border is $y \equiv e^{-\lambda} = e^{\nu} = \frac{1}{k^2 r^2} \sin^2(kr)$, at the border one has $y = 0$, whereas beyond the border surface ($r > r_b$) the well known Reissner-Nordstrøm metric $y(r) = 1 - \frac{2M}{r} + \frac{Q^2}{r^2}$ is valid. It is shown that $M = |Q| = r_b$, so that the exterior metric may also be written as $y = \left(1 - \frac{M}{r}\right)^2$; $r > r_b$ and there is no black hole surrounding the particle.

It is rather remarkable that there exist the considered analytic solution, and it is proposed that this be taken as describing models of classical charged fundamental particles.

The particles presented in [10] and in the present paper are to be regarded as the constituents of elementary particles (like quarks and leptons) and are characterized by their charge being $0$; $\pm\frac{1}{3}e$, with $e$ - the electron charge, as well by radius and mass. It is assumed that every quark or lepton is made up of three of these particles. In a previous paper [10] the relation $M_{neutral} = \frac{1}{2} r_b$ for neutral particles was obtained. Let us take $|Q| = \frac{1}{3} e$ with $e$ being the electron charge. We then have for charged particles $|Q| = M = r_b = 4.59 \times 10^{-35} cm$ and assuming that the neutral one has the same mass as the charged particle, we get $M_{neutral} = 4.59 \times 10^{-35} cm$ ; $r_{b\ neutral} = 9.18 \times 10^{-35} cm$.

Concluding this discussion we want to note the following. If one calculates the matter density in the center of the particle, he obtains $8\pi\rho(0) = k^2 = 4.6846200 \times 10^{69} cm^{-2}$. Comparing this with the Planck density $8\pi\rho_{Pl} = 9.626 \times 10^{66} cm^2$ we see that in the center the substance is beyond the Planck state.

## 7, REFERENCES

## APPENDIX

The metric tensors as given in EQ-s (13), (14) are

$$h_{00} = e^{\nu}; \quad h_{11} = -e^{\lambda}; \quad h_{22} = -r^2; \quad h_{33} = -r^2 \sin^2 \vartheta \qquad (A\text{-}1)$$

$$g_{00} = e^{\tilde{N}(r,l)} \equiv e^{N(l)} h_{00}; \quad g_{11} = -e^{\tilde{L}(r,l)} = e^{L(l)} h_{11}; \quad g_{22} = h_{22}; \quad g_{33} = h_{33}; \quad g_{44} = \varepsilon\, e^{\tilde{F}} \quad (A\text{-}2)$$

For the models considered in Sec.4, Sec5, one has $g_{\mu\nu} = h_{\mu\nu}$.

The basis that accords to (A-1, -2) may be written as

$$e_0^A = e^{-\frac{1}{2}N}, \ 0, \ 0, \ 0, 0. \qquad e_A^0 = e^{\frac{1}{2}N}, \ 0, \ 0, \ 0, 0.$$

$$e_1^A = 0, \ e^{-\frac{1}{2}L}, \ 0, \ 0, 0. \qquad e_A^1 = 0, \ e^{\frac{1}{2}L}, 0, \ 0, 0.$$

$$e_2^A = 0, \ 0, \ 1, 0, 0. \qquad e_A^2 = 0, \ 0, 1, \ 0, \ 0. \qquad \text{(A-3)}$$

$$e_3^A = 0, \ 0, \ 0, \ 1, 0. \qquad e_A^3 = 0, \ 0, \ 0, 1, \ 0.$$

$$n_A = 0, \ 0, \ 0, \ 0, \ \varepsilon \, e^{\frac{1}{2}\tilde{F}}. \qquad n^A = 0, \ 0, \ 0, \ 0, \ e^{-\frac{1}{2}\tilde{F}}.$$

Denoting a partial derivative with respect to $r$ by a prime and that with respect to $l$ by a dot, and taking into account the $r, l$ separation (cf.(15)) we can rewrite the 5D Christoffel symbols

$$\tilde{\Gamma}_{01}^0 = \frac{1}{2}\nu'; \ \tilde{\Gamma}_{04}^0 = \frac{1}{2}\dot{N}; \ \tilde{\Gamma}_{00}^1 = \frac{1}{2}e^{\tilde{N}-\tilde{L}}\nu'; \ \tilde{\Gamma}_{11}^1 = \frac{1}{2}\lambda'; \ \tilde{\Gamma}_{14}^1 = \frac{1}{2}\dot{L}; \ \tilde{\Gamma}_{22}^1 = -re^{-\tilde{L}};$$

$$\tilde{\Gamma}_{33}^1 = -r\sin^2\vartheta \, e^{-\tilde{L}}; \ \tilde{\Gamma}_{44}^1 = \frac{\varepsilon}{2}e^{\tilde{F}-\tilde{L}}\psi'; \ \tilde{\Gamma}_{12}^2 = \frac{1}{r}; \ \tilde{\Gamma}_{33}^2 = -\sin\vartheta\cos\vartheta;$$

$$\tilde{\Gamma}_{13}^3 = \frac{1}{r}; \ \tilde{\Gamma}_{23}^3 = \cot\vartheta; \ \tilde{\Gamma}_{00}^4 = -\frac{\varepsilon}{2}e^{\tilde{N}-\tilde{F}}\dot{N}; \ \tilde{\Gamma}_{11}^4 = \frac{\varepsilon}{2}e^{\tilde{L}-\tilde{F}}\dot{L}; \qquad \text{(A-4)}$$

$$\tilde{\Gamma}_{14}^4 = \frac{1}{2}\psi'; \ \tilde{\Gamma}_{44}^4 = \frac{1}{2}\dot{F};$$

and the 4D Christoffel symbols

$$\Gamma_{01}^0 = \frac{1}{2}\nu'; \ \Gamma_{00}^1 = \frac{1}{2}e^{\nu-\lambda}\nu'; \ \Gamma_{11}^1 = \frac{1}{2}\lambda'; \ \Gamma_{22}^1 = -r\,e^{-\lambda};$$

$$\Gamma_{33}^1 = -r\,e^{-\lambda}\sin^2\vartheta; \ \Gamma_{12}^2 = \frac{1}{r}; \ \Gamma_{33}^2 = -\sin\vartheta\cos\vartheta; \ \Gamma_{13}^3 = \frac{1}{r}; \ \Gamma_{23}^3 = \cot\vartheta; \qquad \text{(A-5)}$$

Making use of EQ. (16-18) as well of (A-3) – (A-5) one obtains for the quantities appearing in (20) and listed in (12a-12e)

$$M_0^0 = M_1^1 = -M_2^2 = -M_3^3 = \frac{1}{8\pi}e^{-(\lambda+\nu)}(w_0')^2 \qquad \text{(A-6a)}$$

$$B_{01} \equiv B_{10} = -e^{-\frac{1}{2}(L+N+2\tilde{F})}\dot{\tilde{w}}_0 \tilde{w}_{4,1}; \quad B_0^0 = e^{-(\nu+\psi)}(\dot{\tilde{w}}_0)^2;$$

$$B_1^1 = -e^{-(\lambda+\psi)}(\tilde{w}_{4,1})^2; \quad B = e^{-(\nu+\psi)}(\dot{\tilde{w}}_0)^2 - e^{-(\lambda+\psi)}(\tilde{w}_{4,1})^2 \qquad \text{(A-6b)}$$

$$C_{00} = \frac{1}{2}e^{\nu-\frac{1}{2}\psi}\dot{N}; \quad C_{11} = -\frac{1}{2}e^{\lambda-\frac{1}{2}\psi}\dot{L} \qquad \text{(A-6c)}$$

$$E_0^0 = \frac{\varepsilon}{4} e^{-\lambda} \nu' \psi' - \frac{1}{2} e^{-\psi} \left[ \ddot{N} + \frac{1}{2} (\dot{N})^2 - \frac{1}{2} \dot{N} \dot{F} \right]$$

$$E_1^1 = \frac{\varepsilon}{2} e^{-\lambda} \left[ \psi'' + \frac{1}{2} (\psi')^2 - \frac{1}{2} \lambda' \psi' \right] - \frac{1}{2} e^{-\psi} \left[ \ddot{L} + \frac{1}{2} (\dot{L})^2 - \frac{1}{2} \dot{L} \dot{F} \right] \quad \text{(A-6d)}$$

$$E_2^2 = E_3^3 = \frac{\varepsilon}{2} e^{-\lambda} \frac{\psi'}{r}; \quad E_{01} = 0 .$$

$$E \equiv E_\sigma^\sigma = \frac{\varepsilon}{2} e^{-\lambda} \left[ \psi'' + \frac{1}{2} (\psi')^2 + \frac{1}{2} \psi'(\nu' - \lambda') + 2 \frac{\psi'}{r} \right]$$
$$- \frac{1}{2} e^{-\psi} \left[ \ddot{N} + \ddot{L} + \frac{1}{2} (\dot{N})^2 + \frac{1}{2} (\dot{L})^2 - \frac{1}{2} \dot{F} (\dot{N} + \dot{L}) \right] \quad \text{(A-6e)}$$